\documentclass[aps,epsf,nofootinbib, notitlepage,showpacs,11pt, unsortedaddress, superscriptaddress]{revtex4-2}
\pdfoutput=1 
\usepackage{graphicx} 
\PassOptionsToPackage{square,numbers,sort&compress}{natbib}
\usepackage{fontawesome}
\usepackage[utf8]{inputenc} 
\usepackage[T1]{fontenc}    
\usepackage{hyperref}       
\hypersetup{
    colorlinks=true,        
    linkcolor=black,        
    citecolor=blue,         
    urlcolor=blue,          
    filecolor=blue          
}
\usepackage{url}            
\usepackage{booktabs}       
\usepackage{algorithm}      
\usepackage{amsmath, amsfonts}       
\usepackage{nicefrac}       
\usepackage{svg}            
\usepackage{microtype}      
\usepackage{multirow}       
\usepackage[table,x11names]{xcolor}
\usepackage{multirow}       
\usepackage{float}          
\usepackage{appendix}       
\usepackage{setspace}       

\usepackage[hang,flushmargin]{footmisc}
\usepackage{xcolor}
\definecolor{lightergray}{gray}{0.9}

\usepackage{array}

\usepackage{pgfplots}
\pgfplotsset{width=10cm,compat=1.9}

\usepackage{tikz}

\usepackage{algorithm}
\usepackage{algpseudocode}

\begin{document}

\title{\Large{SpinGQE: A generative quantum eigensolver for spin Hamiltonians}}
\author{Alexander Holden}
\thanks{Work done at Mindbeam AI}
\affiliation{Mindbeam AI}
\email{research@mindbeam.ai}
\affiliation{Department of Computer and Data Sciences, Case Western Reserve University}
\author{Moinul Hossain Rahat}
\author{Nii Osae Osae Dade}
\affiliation{Mindbeam AI}

\begin{abstract}
The ground state search problem is central to quantum computing, with applications spanning quantum chemistry, condensed matter physics, and optimization. The Variational Quantum Eigensolver (VQE) has shown promise for small systems but faces significant limitations. These include barren plateaus, restricted ansatz expressivity, and reliance on domain-specific structure. We present SpinGQE, an extension of the Generative Quantum Eigensolver (GQE) framework to spin Hamiltonians. Our approach reframes circuit design as a generative modeling task. We employ a transformer-based decoder to learn distributions over quantum circuits that produce low-energy states. Training is guided by a weighted mean-squared error loss between model logits and circuit energies evaluated at each gate subsequence. We validate our method on the four-qubit Heisenberg model, demonstrating successful convergence to near-ground states. Through systematic hyperparameter exploration, we identify optimal configurations: smaller model architectures (12 layers, 8 attention heads), longer sequence lengths (12 gates), and carefully chosen operator pools yield the most reliable convergence. Our results show that generative approaches can effectively navigate complex energy landscapes without relying on problem-specific symmetries or structure. This provides a scalable alternative to traditional variational methods for general quantum systems. An open-source implementation is available at \faGithub~\url{https://github.com/Mindbeam-AI/SpinGQE}.
\end{abstract}

\maketitle

\section{Introduction}
Finding the ground state of a quantum system is a foundational task in quantum computing. Applications range from quantum chemistry and condensed matter physics to combinatorial optimization and quantum machine learning \cite{Bauer:2020epk, McArdle:2018tza}. Crucial physical properties are encoded within the ground state, such as the minimum energy configuration of a molecule or the solution to a spin system. This makes the development of accurate and scalable ground state solvers a key challenge in quantum computing.

The Variational Quantum Eigensolver (VQE) \cite{Peruzzo:2013bzg, Cerezo:2020jpv} is among the leading approaches for noisy intermediate-scale quantum (NISQ)-era devices \cite{Preskill:2018jim, Bharti:2021zez}. VQE optimizes a parameterized quantum circuit, or ansatz, to minimize the expectation value of a given Hamiltonian. It has demonstrated success for small molecules and structured Hamiltonians but faces challenges as system size and complexity grow. Barren plateaus in the optimization landscape, limited expressivity of ansatzes, and sensitivity to circuit depth and noise are significant sources of error \cite{McClean:2018jps, Wang:2020yjh}. Moreover, VQE methods rely on physical intuition and domain knowledge, such as symmetries in electronic structure problems. These do not generalize to arbitrary Hamiltonians.

To address these limitations, recent work has proposed the Generative Quantum Eigensolver (GQE) \cite{Nakaji:2024oyd}, a novel method that reframes circuit design as a generative modeling task. Instead of optimizing continuous parameters within a fixed circuit, GQE uses a classical generative model to learn a distribution over full quantum circuits that produce low-energy states. This shifts the optimization burden to the classical side \cite{Liang:2023ptk, jaouni2024deep}. Specifically, we employ an approach where a transformer model is trained to predict circuit energies using an online training loop. This enables scalable training using standard deep learning tools while still leveraging quantum devices to evaluate circuit energies.

In this work, we introduce SpinGQE, an extension of the Generative Quantum Eigensolver (GQE) framework to spin Hamiltonians. While the original GQE was developed for fermionic systems \cite{Nakaji:2024oyd}, spin Hamiltonians represent a fundamentally different class of quantum systems with direct applications in combinatorial optimization \cite{Lucas:2013ahy}, magnetic material design \cite{Zahedinejad:2017kvw}, and quantum state engineering. By adapting the generative circuit modeling approach to the native Pauli structure of spin systems, we demonstrate that GQE-based methods can serve as versatile ground state solvers across diverse quantum domains. Our approach maintains the core advantages of the GQE framework, i.e., avoiding barren plateaus and restrictive ansatz design, while introducing domain-specific innovations including weighted loss functions and hybrid classical-quantum post-processing optimization.

By extending GQE to spin Hamiltonians and validating its performance in this new regime, we aim to push beyond the limitations of existing variational methods. Our goal is to move toward more general-purpose, scalable quantum eigensolvers.

\section{Methods}
\subsection{Generative quantum eigensolver framework}
Our approach builds on the Generative Quantum Eigensolver framework \cite{nakaji2024generative}, which we have adapted for spin Hamiltonians. The implementation is based on the open-source PennyLane framework \cite{bergholm2018pennylane}, extending the methodology presented in their GQE tutorial \cite{bunao2024gqe}. The model is trained using an online generative approach. A transformer-based decoder outputs quantum circuits as sequences of tokens sampled from a fixed operator pool.

At each data generation step, we sample a batch of operator sequences, or circuits, from the model's current distribution. For each operator sequence $U = (U_0 U_1 \ldots U_N)$, we compute the expectation value of the target Hamiltonian $H$ on a quantum computer or simulator:
\begin{equation}
E(H, \psi) = \langle \psi | H | \psi \rangle,
\end{equation}
where $\psi$ is the quantum state obtained from the quantum circuit $U$. 

Rather than evaluating only the final circuit energy, we evaluate the energy of every prefix subsequence of gates. For a sequence $U$ of length $N$, we execute $U_{1:t} = (U_0 U_1 \ldots U_t)$ for all $t = 1, 2, \ldots, N$. This produces a trajectory of energies through the circuit. Concurrently, the model produces a series of unnormalized cumulative logits, one for each timestep. These reflect the model's internal ``score'' for generating that prefix. 

We train the model to match these cumulative logits to the energies of the corresponding circuits. We apply a mean-squared error (MSE) loss between the cumulative logits and measured energies:
\begin{equation}
\text{MSE} = \frac{1}{N} \sum_{i=1}^{N} \sum_{t=1}^{T^{(i)}} \left( l_t(x^{(i)}) - E_t^{(i)} \right)^2,
\end{equation}
where $l_t(x^{(i)})$ is the cumulative logit at step $t$ for the $i$-th circuit sequence, and $E_t^{(i)}$ is the corresponding energy of the subsequence up to that point. This per-subsequence supervision encourages the model to learn progressive energy descent patterns. It also makes it easier for the model to assign credit to effective partial circuits.

\begin{figure}
    \centering
    \includegraphics[width=1.02\linewidth]{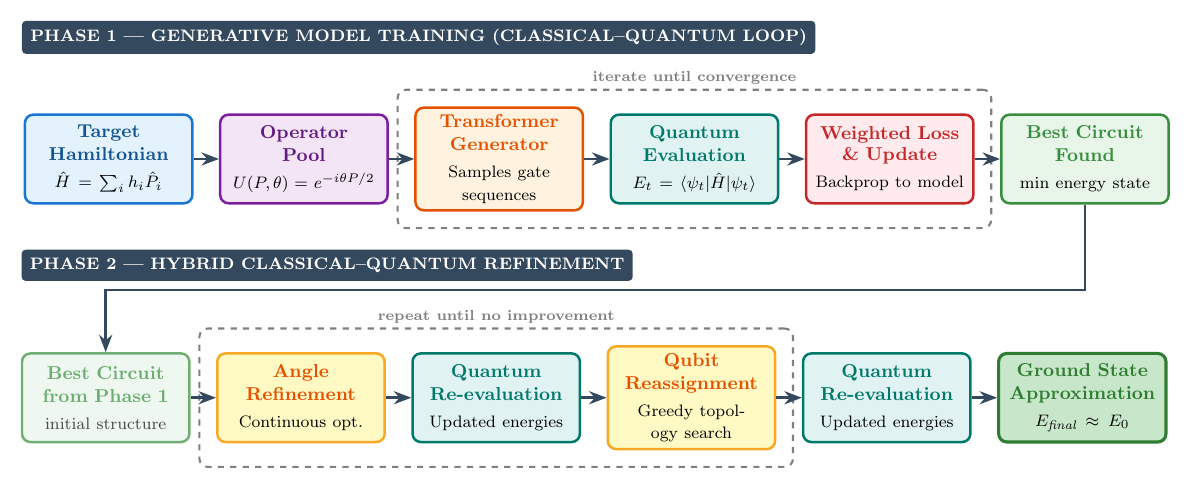}
    \caption{SpinGQE methodology for Hamiltonian ground state search. The framework iteratively samples quantum circuits from a transformer model, evaluates their energies on a quantum device, and updates the model to favor low-energy circuits through weighted loss optimization. An optimization loop can further refine the circuits to lower the energy.}
    \label{fig:flowchart}
\end{figure}

\subsection{Weighted loss function}
We employ a weighted loss function to prioritize accuracy in low-energy regions. The weights are applied according to:
\begin{equation}
w(E, \beta) = \frac{1}{1 + e^{\beta E}}.
\end{equation}
This gives higher weight to circuits with lower energies. The weighted MSE loss becomes:
\begin{equation}
\text{MSE}_{\text{weighted}} = \frac{1}{N} \sum_{i=1}^{N} w(E_N^{(i)}) \sum_{t=1}^{T^{(i)}} \left( l_t(x^{(i)}) - E_t^{(i)} \right)^2,
\end{equation}
where $E_N^{(i)}$ denotes the final energy of the complete circuit. The parameter $\beta$ controls the sharpness of the weighting function. Higher $\beta$ values focus on near-zero energy circuits, while lower values provide more uniform weighting across the energy range. This discourages premature convergence to local minima.

\subsection{Operator pool design} \label{sec:2.3}
For spin Hamiltonians, we construct operator pools that reflect the physical structure of the target system. Our selected operator pool mirrors the Pauli rotations in the Hamiltonian. Specifically, we use a set of Pauli rotations:
\begin{equation}
U(P, \theta) = e^{-i\frac{\theta}{2} P},
\end{equation}
where $P$ is a Pauli string from the set of Hamiltonian terms $\{Z, XX, YY, ZZ\}$ and $\theta$ is an angle chosen from the discrete set $\{\pm \pi/2^k \mid k = 1, \ldots, 5\}$. For single-qubit rotations, we include operators acting on all individual qubits. For two-qubit rotations, we restrict to nearest-neighbor pairs $\{i, i+1\}$, respecting the one-dimensional structure of the Heisenberg chain. This design balances expressivity with computational tractability. It provides sufficient flexibility to explore the relevant portion of Hilbert space while maintaining a manageable discrete search space. The effect of extending this operator pool is discussed further in appendix~\ref{app:C}.

\subsection{Post-processing optimization}
After training the model using the methodology described above, we implement a post-processing optimization pipeline. This combines classical angle refinement (L-BFGS-B \cite{byrd1995limited} or COBYLA \cite{powell1994direct}) with a greedy search over qubit pairs on which the two-qubit operators act. This post-processing refines the model's initial circuit by dynamically adjusting gate placement and rotation angles to minimize energy further. By doing so, we enable circuits to escape the constraints imposed by the discrete operator pool. This allows for long-range entanglement and adaptive rotation angles. We found this hybrid approach superior to training the model on a larger operator pool. It maintains a manageable search space while achieving high precision.

\begin{algorithm}[H]
    \caption{Post-Processing Optimization Loop}
    \label{alg:myalgorithm}
    \begin{algorithmic}[1]
        \State $\text{sequence, energy} \gets \text{angleRefinement(sequence)}$
        \For {$i$ in range(len(sequence))}
            \State $\text{op} \gets \text{sequence}[i]$
            \State $\text{trialQubits} \gets \text{combinations(allQubits, len(op.qubits))}$
            \For {$q$ in trialQubits}
                \State $\text{tempSequence} \gets \text{sequence}$
                \State $\text{tempSequence}[i] \gets \text{newOp(op.type, op.angle, } q\text{)}$
                \State $\text{tempSequence, tempEnergy} \gets \text{angleRefinement(tempSequence)}$
                \If {$\text{tempEnergy} < \text{energy}$}
                    \State $\text{sequence, energy} \gets \text{tempSequence, tempEnergy}$
                \EndIf
            \EndFor
        \EndFor
        \State \textbf{return} sequence, energy
    \end{algorithmic}
\end{algorithm}

\section{Results and discussion}
\subsection{Target system: one-dimensional Heisenberg model}
We validate our approach on the four-qubit one-dimensional Heisenberg model with an external magnetic field \cite{auerbach1994interacting, sachdev2011quantum}, described by the Hamiltonian:
\begin{equation}
H = J\sum_{n=1}^{N-1} (X_n X_{n+1} + Y_n Y_{n+1} + Z_n Z_{n+1}) + h \sum_{n=1}^{N} Z_n,
\end{equation}
where $J$ represents the isotropic exchange coupling, $h$ denotes the external magnetic field strength, and $N = 4$ is the system size. This model exhibits distinct physical regimes depending on the relative magnitudes of $J$ and $h$, providing a rigorous test of our method's ability to adapt to different quantum phases.

In the antiferromagnetic regime ($h \leq J$), exchange interactions dominate, leading to frustrated, highly entangled ground states with significant quantum fluctuations. The energy landscape features numerous local minima and extended plateaus, presenting substantial optimization challenges. Conversely, in the field-dominated regime ($h > J$), the external field aligns spins, producing ground states that approach product states with reduced entanglement and smoother energy landscapes.

\subsection{Model configuration and hyperparameters}

Through systematic exploration of the hyperparameter space, we identify an optimal transformer configuration comprising 12 hidden layers and 8 attention heads with an embedding dimension of 512, totaling 37.83 million parameters. This architecture balances model capacity with training stability, proving more effective than larger configurations which exhibited premature convergence to local minima.

Our operator pool mirrors the Pauli structure of the Hamiltonian as described in Section \ref{sec:2.3}, with discrete angles $\{\pm \pi/2^k \mid k = 1, \ldots, 5\}$. We sample circuits of length 12 gates, applying logit-sharpening with generation temperature $\tau = 0.5$ and energy weighting parameter $\beta = 0.3$. We use an AdamW optimizer with learning rate $4 \times 10^{-4}$. Training proceeds for 700 epochs with $M = 10$ circuits generated per epoch, providing sufficient exploration of the circuit space while maintaining computational efficiency. We save checkpoints at every 50 epochs. We average the best three checkpoints based on the minimum energy of the generated circuits calculated using a Quantum Processing Unit (QPU) to determine the final model.

The hyperparameter choices reflect careful trade-offs specific to spin systems. Smaller models prevent overfitting to spurious local minima in the rugged Heisenberg energy landscape. Moderate sequence length (12 gates) provides sufficient expressivity without excessive circuit depth that could complicate training. The selected $\beta$ value balances exploration of the energy landscape with sufficient focus on low-energy regions.

In Appendices \ref{app:A} and \ref{app:D}, we explore the effect of several model sizes and the choice of the parameters $\beta$ and $M$ on model convergence and estimating the ground state energy for the case of $h=J=10$.

\subsection{Antiferromagnetic regime ($h \leq J$) results}

We first examine the challenging antiferromagnetic regime where $h \leq J$. For concreteness, we focus on the case $h = J = 10$, where competing exchange interactions produce a highly frustrated ground state. The exact ground state energy for this configuration is $E_0^{\rm exact} = -64.641J$. This parameter choice tests the model's ability to navigate complex energy landscapes characterized by quantum frustration and high entanglement.

\begin{figure}[htbp]
\centering
\scalebox{0.90}{
\includegraphics[width=0.8\textwidth]{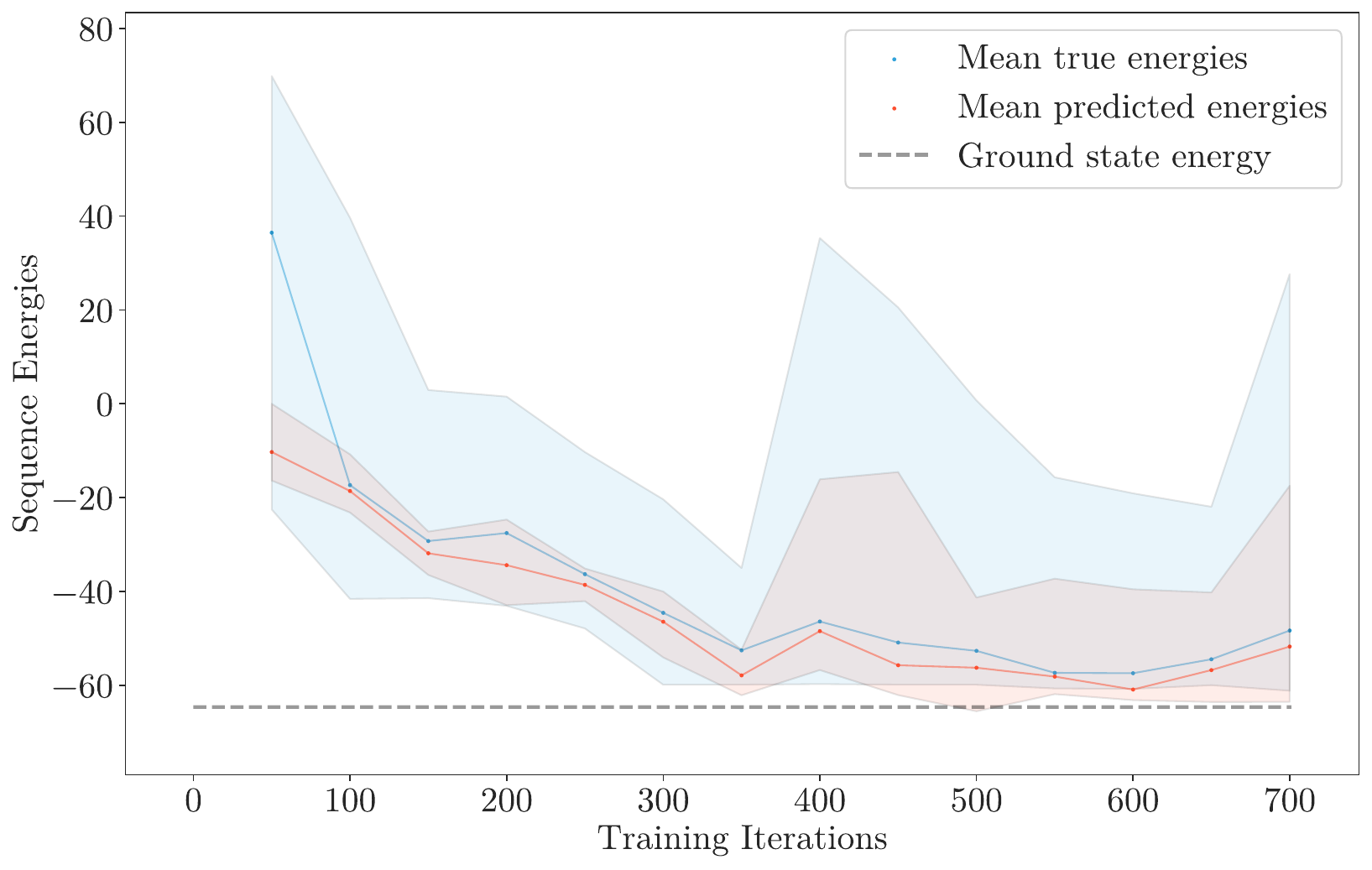}
}
\caption{Training convergence of SpinGQE on the Heisenberg Hamiltonian with $J=h=10$. The plot shows the energy achieved by generated circuits as a function of training epoch. Model size is 37M parameters, with hyperparameters $\beta = 0.3$ and $M=10$ circuits per epoch. The shaded regions correspond to the range of energies of these $10$ circuits, while the solid line represents their average.}
\label{fig:Fig1}
\end{figure}

Prior to post-processing optimization, the model successfully generates circuits with energies as low as $E_0^{\rm model} = -60.78 J$ after approximately 350 training epochs, as shown in Fig.~\ref{fig:Fig1}. The convergence trajectory exhibits initial rapid descent followed by gradual refinement. This is characteristic of effective exploration transitioning to exploitation. It indicates that the weighted loss function successfully guides the model toward low-energy circuits.

We select the best three checkpoints based on the average energy of the generated circuits. We average these checkpoints to obtain the final model.

\begin{figure}
    \centering
    \includegraphics[width=\linewidth]{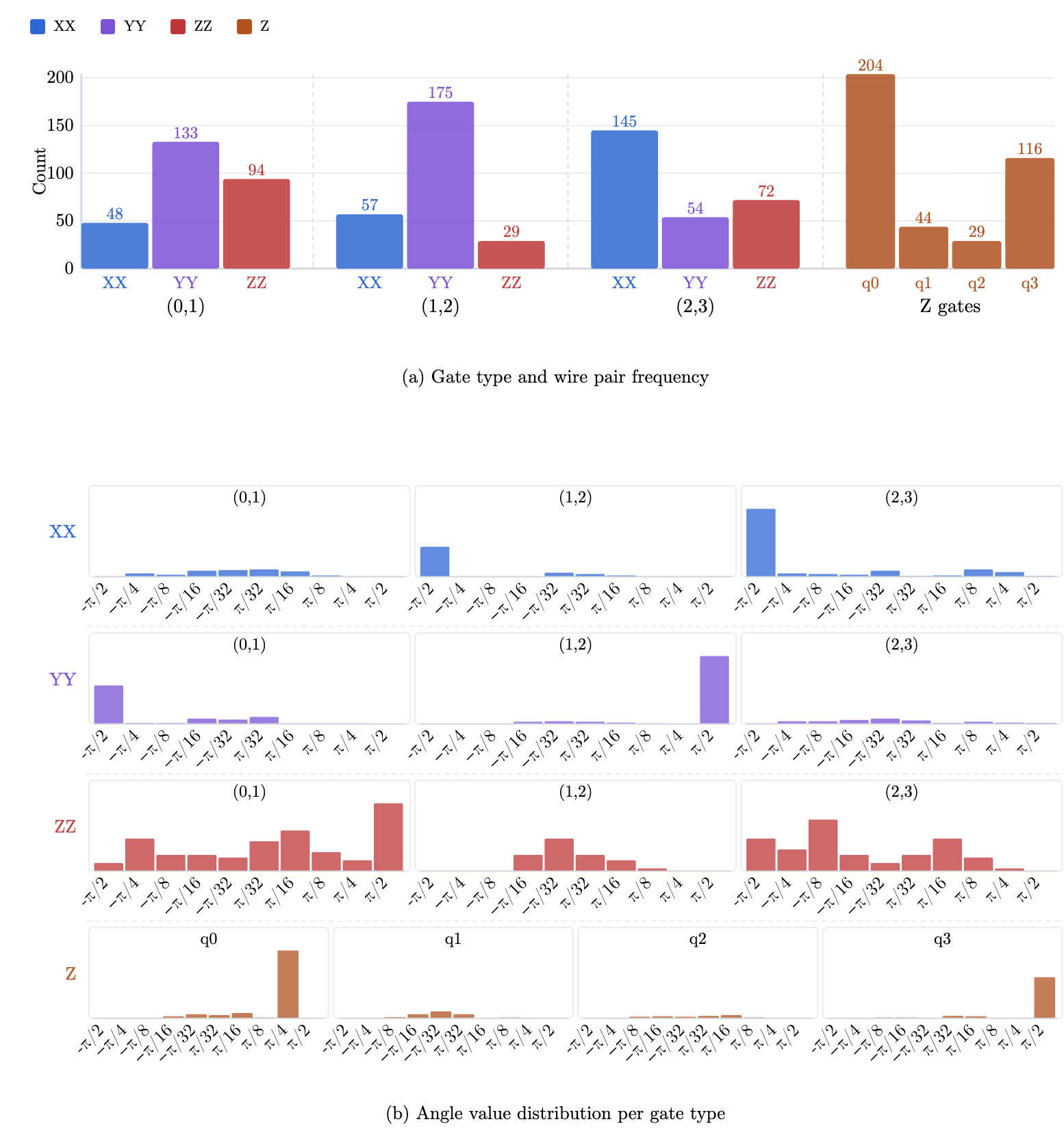}
    \caption{Learned gate and angle distributions from the final model. (a) Frequency of gate types and qubit pair combinations across 100 sampled circuits. (b) Distribution of rotation angles for each gate type and qubit pair, showing the model's learned preferences for specific parameter values.}
    \label{fig:distribution}
\end{figure}

To gain further insight into the types of gates that appear in sampled circuits from the final model, we show the distribution of 100 sampled circuits. Figure~\ref{fig:distribution} displays the distribution by gate type, wire pair for entangling gates, and angle values. The base model learns to generate quantum circuits by sampling from a discrete operator pool. This pool consists of two-qubit Pauli rotation gates (XX, YY, ZZ) on nearest-neighbor qubit pairs and single-qubit Z rotations.

Figure~\ref{fig:distribution}(a) shows the frequency with which each gate type and qubit pair combination is generated across 100 sampled circuits. The model exhibits clear preferences for specific gate and wire pair combinations rather than generating them uniformly. YY gates dominate on the $(1,2)$ pair with 175 occurrences, and XX gates are most frequent on the $(2,3)$ pair with 145 occurrences. Among single-qubit gates, Z rotations on qubit 0 appear far more frequently than on other qubits, with 204 occurrences compared to 44, 29, and 116 for qubits 1, 2, and 3 respectively. Figure~\ref{fig:distribution}(b) reveals that the angle distributions vary considerably across gate type and qubit pair combinations. Notably, YY gates on the $(1,2)$ pair are strongly biased toward $+\pi/2$, accounting for 155 out of 175 occurrences. Similarly, YY gates on the $(0,1)$ pair and XX gates on the $(2,3)$ pair are heavily concentrated at $-\pi/2$, with 88 and 102 occurrences respectively. Z gates on qubit 0 show a pronounced preference for $+\pi/4$ with 165 occurrences, while Z gates on qubit 3 are concentrated at $+\pi/2$ with 100 occurrences. This suggests the model has identified specific gate-angle combinations that are particularly effective for this Hamiltonian. Other combinations such as ZZ gates and Z rotations on qubits 1 and 2 exhibit flatter angle distributions, indicating the model has not developed strong preferences for those parameters.

\begin{figure}
    \centering
    \includegraphics[width=\linewidth]{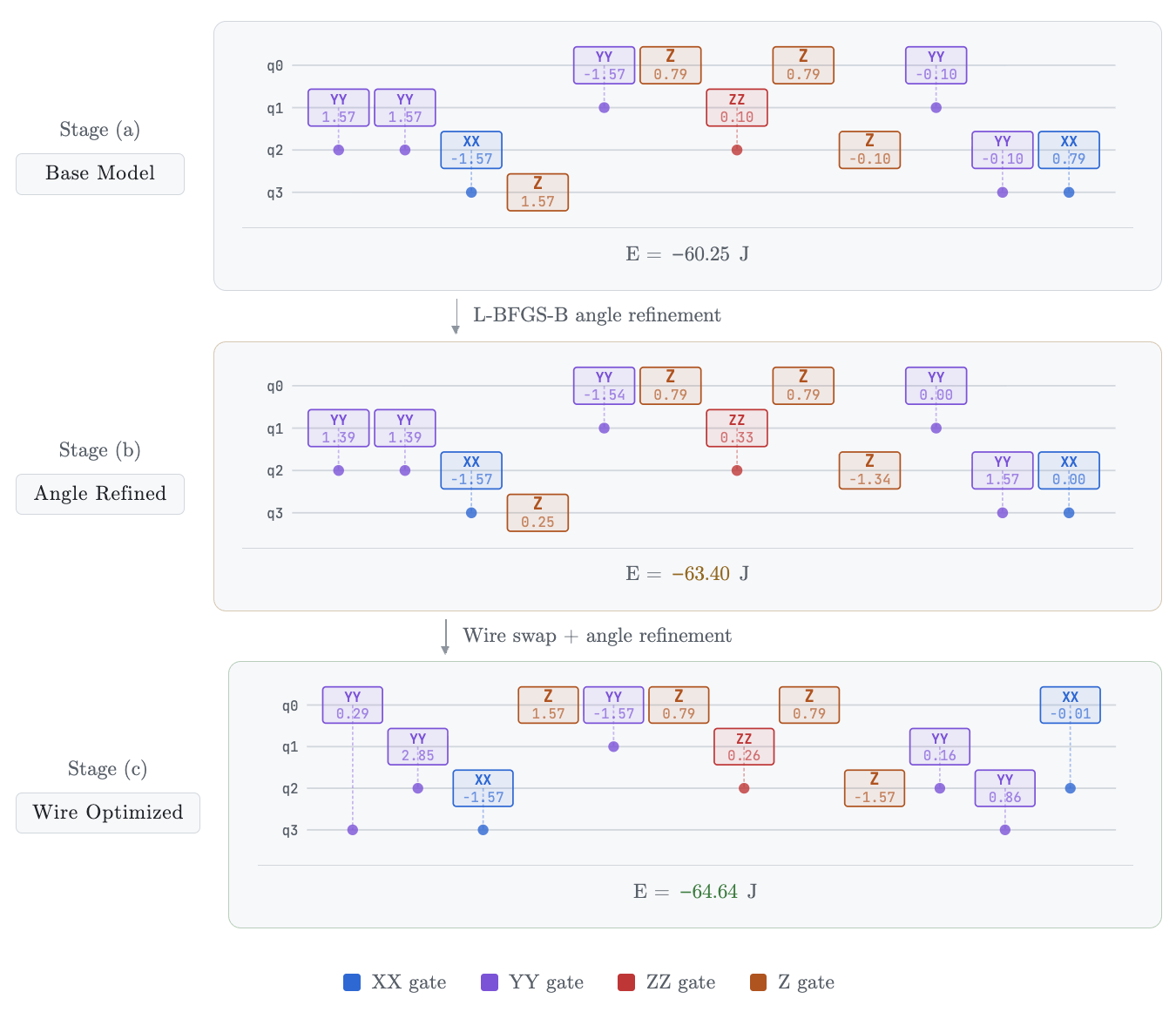}
    \caption{Progressive optimization of a generated circuit through post-processing stages. The base model output (left) undergoes angle refinement (middle) and wire swap optimization (right), achieving successive energy reductions to reach near-exact ground state energy. Circuit diagrams show the gate sequences at each stage.}
    \label{fig:optimization}
\end{figure}

Figure~\ref{fig:optimization} illustrates the progressive energy optimization applied to a circuit sampled from the base model. The base model generates a circuit with energy $-60.25 J$, already capturing substantial ground state structure through learned gate selection. Angle refinement using L-BFGS-B optimization of the Pauli rotation angles then reduces the energy to
$-63.40 J$. This demonstrates that the discrete angle values used during training are suboptimal, and that continuous optimization can extract additional energy from the same circuit topology. The subsequent wire swap stage greedily searches over alternative qubit assignments for each gate, re-optimizing angles after each change. This further reduces the energy to
$-64.64 J$, essentially reaching the exact ground state energy of
$-64.641 J$ obtained by exact diagonalization. The wire swap stage is particularly important because the base model is constrained to generate only nearest-neighbor gates, whereas the optimal circuit requires long-range entanglement between non-adjacent qubits such as
$(0,3)$ and
$(1,2)$. Together, the two post-processing stages improve the circuit energy by
$4.39 J$. Angle refinement contributes approximately $70\%$, and wire reassignment contributes the remaining $30\%$.

\subsection{Field-dominated regime ($h>J$) results}

\begin{figure}[htbp]
\centering
\scalebox{0.90}{
\includegraphics[width=0.8\textwidth]{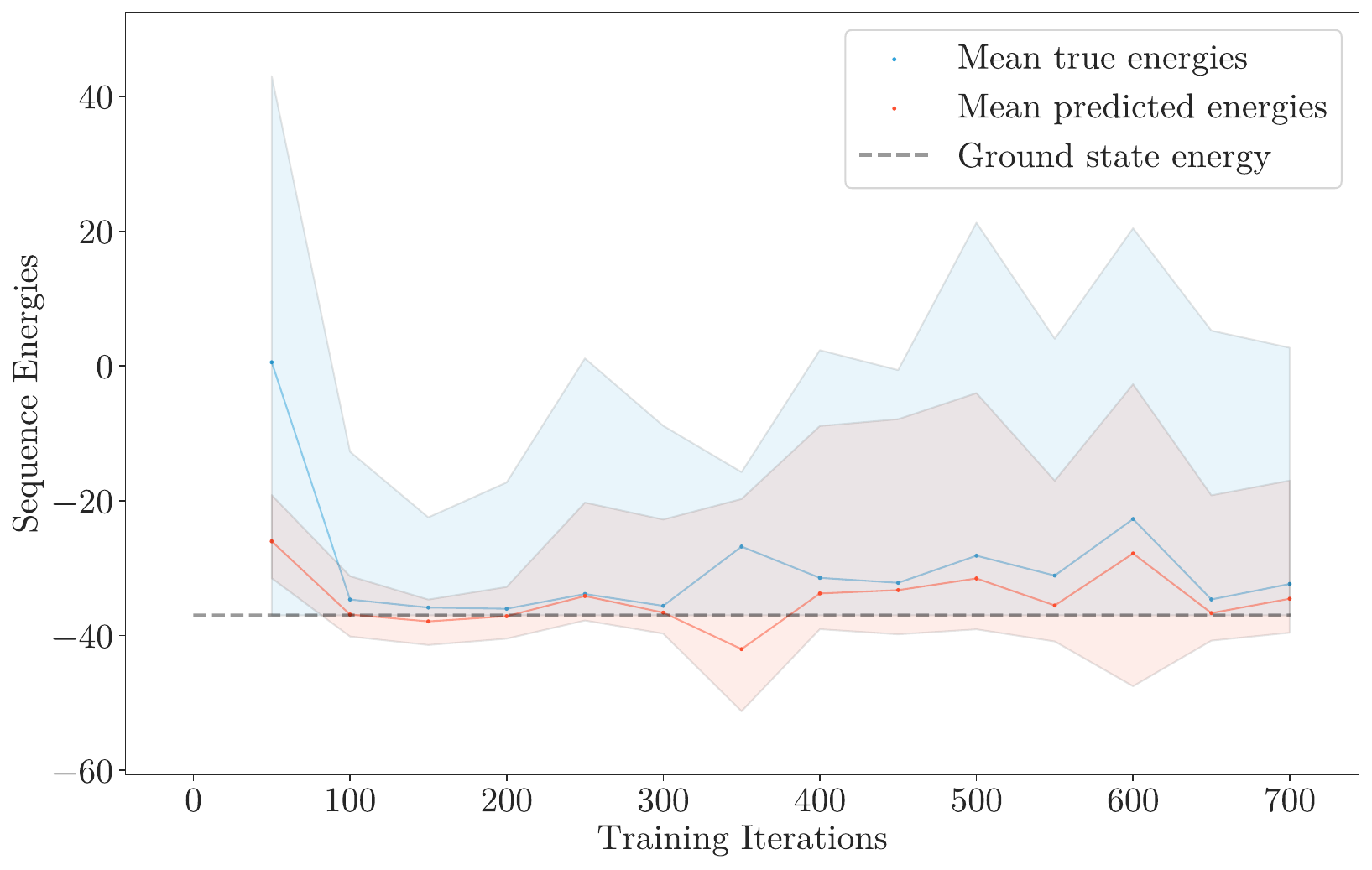}
}
\caption{Training convergence of SpinGQE on the Heisenberg model with $J=1, h=10$. The field-dominated regime exhibits smoother convergence to the ground state compared to the antiferromagnetic regime, reflecting the simpler energy landscape.}
\label{fig:Fig2}
\end{figure}

To test our method in this alternative energy setting, we chose parameters $J=1$ and $h=10$. This Hamiltonian differs from the previous case, since the external magnetic field aligns the spins in its direction. This creates a computationally easier scenario, as the Hamiltonian approaches a product state. Using the same methods, after model convergence, the model produces circuits with energy $-37.0 J$. This already approximates the ground state well and thus does not require the post-processing loop, as seen in Fig.~\ref{fig:Fig2}. Although this plot corresponds to the case $\beta=0.3$ and $M=10$, a grid search similar to the $h \leq J$ case shows that the true ground state is achieved in all combinations of $(\beta, M)$ in the field-dominated regime.

\begin{figure}[htbp]
\centering
\scalebox{1.0}{
\includegraphics[width=\textwidth]{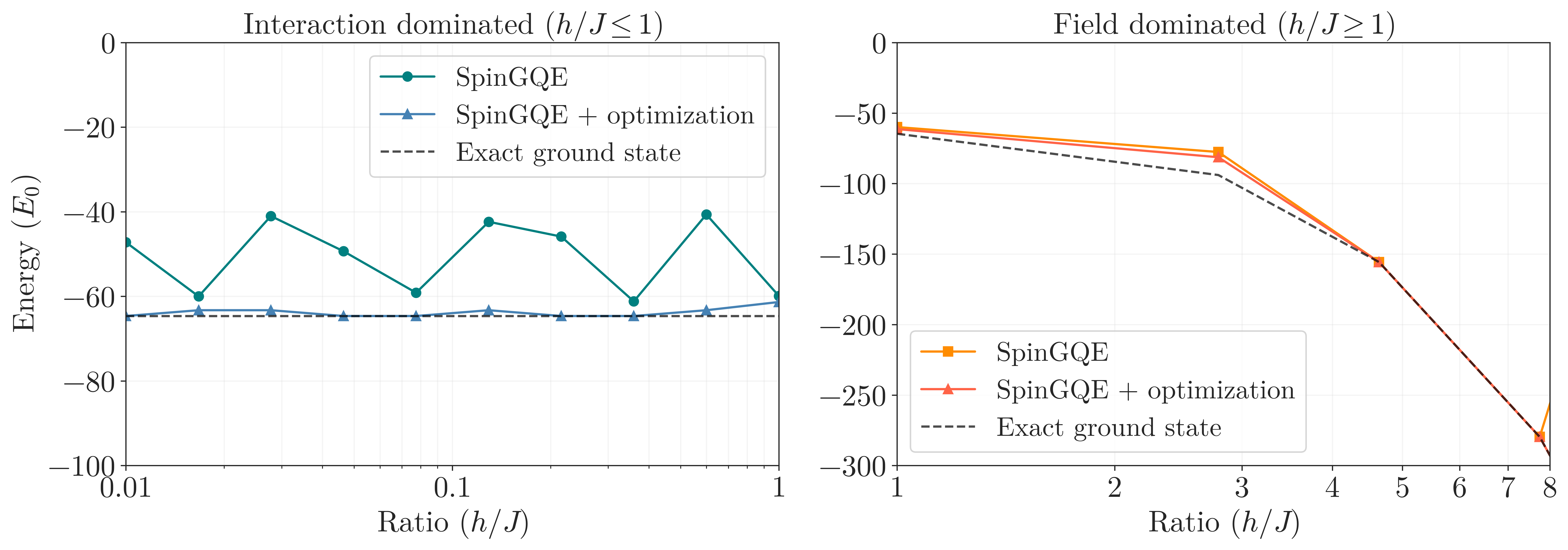}
}
\caption{Ground state energy estimation across different physical regimes of the Heisenberg model. The SpinGQE method with the optimization loop successfully identifies near-ground states for both antiferromagnetic ($h/J \leq 1$) and field-dominated ($h/J > 1$) regimes, demonstrating robustness across different energy landscape topologies.}
\label{fig:regime}
\end{figure}

Following the above procedure, we scan over the ratio $h/J$ in both regimes, namely $0.01 \leq h/J \leq 1$ and $1 \leq h/J \leq 8$. The algorithm generates good approximations to the true ground state energy in both cases, as shown in Fig.~\ref{fig:regime}. We show the results from the SpinGQE algorithm before and after the optimization loop. 
The optimization loop bridges the gap between the prediction of the SpinGQE algorithm and the true ground state energy for the $h \leq J$ case. For $h \geq J$, the SpinGQE algorithm itself approximates the exact ground state energy closely. 

We observe a remarkable property of the one-dimensional antiferromagnetic Heisenberg Hamiltonian 
for systems with an even number of qubits: the ground state energy remains strictly constant for magnetic field strengths $h \leq J$. This phenomenon arises from the commutation $[H_0, S_z^{\text{total}}] = 0$, where $H_0$ denotes the Heisenberg exchange term and $S_z^{\text{total}}$ is related to the Zeeman term:
\begin{align}
    H_0 &= \sum_{n=1}^{N-1} (X_n X_{n+1} + Y_n Y_{n+1} + Z_n Z_{n+1}), \\
    S_z^{\text{total}} &= \frac{1}{2}\sum_{n=1}^{N} Z_n.
\end{align}
For even $N$, the ground state at $h=0$ is a spin singlet with vanishing total magnetization, $\langle S_z^{\text{total}} \rangle = 0$. Since the Zeeman term is diagonal in the eigenbasis of $H_0$ and the ground state resides in the $S_z^{\text{total}} = 0$ sector, the first-order perturbative correction vanishes identically: $\Delta E^{(1)} = 2h\langle S_z^{\text{total}} \rangle = 0$. Moreover, all higher-order corrections also vanish due to the absence of off-diagonal matrix elements within the $S_z^{\text{total}} = 0$ subspace. Consequently, the ground state remains unchanged until $h$ exceeds a critical field $h_c \approx 1.3J$, at which point a level crossing occurs and the system transitions to a polarized state with $\langle S_z^{\text{total}} \rangle \neq 0$. This behavior exemplifies symmetry protection in quantum many-body systems, where a conserved quantum number shields the ground state from external perturbations \cite{auerbach1994interacting}. Our algorithm is able to approximate this ground state energy across orders of magnitude of $h/J$ in the antiferromagnetic regime.

\section{Conclusion}

In this work, we demonstrated SpinGQE, a new approach to the quantum ground state problem for spin Hamiltonians using the Generative Quantum Eigensolver framework combined with a post-processing optimization loop. By adopting an online training strategy and aligning cumulative model logits to per-subsequence circuit energies, we enable the model to learn meaningful structure from the quantum energy landscape. Additionally, the post-processing loop allows the generated circuits to escape the strict boundaries imposed by discrete operator pools and angles. This approach highlights the potential of circuit-level generation to address the challenges posed by the ground state search problem.

While our results demonstrate the potential of generative models for ground state discovery, several challenges remain. These represent directions for future research. A primary concern is the computational complexity associated with increasing the number of qubits and corresponding circuit length. As the system grows, the Hilbert space expands exponentially. This makes the discovery of refined operator pools and more efficient scaling methods essential.

However, it is notable that the scaling overhead remains largely classical. It centers on the Transformer's ability to process longer and more complex sequences of gate tokens. This suggests a potential advantage over traditional methods, though it does not exclude the possible need for larger training datasets for larger models. Future investigation into transfer learning or multi-task pre-training could mitigate these data requirements. This would allow models trained on smaller spin chains to generalize to the complex, long-range correlations of larger Hamiltonians.

This work demonstrates a new approach to the Hamiltonian ground state search problem, specifically applied to spin Hamiltonians. By aligning circuit generation with energy evaluation through logit-matching, we provide a flexible alternative to traditional variational methods. These approaches could potentially serve as the foundation for more robust eigensolvers that adapt dynamically to diverse problem settings.

\appendix

\section{Data generation and model sizes} \label{app:A}

A core objective of SpinGQE is to maximize ``computational offloading'' by shifting the burden of state discovery from expensive quantum processing unit (QPU) evaluations to classical generative modeling. In this limited-resource regime, the choice of model architecture becomes a trade-off between representational power and sample efficiency. Smaller, more compact Transformers are superior in this context, as they avoid the data requirements typical of larger architectures. Utilizing this streamlined model ensures more stable convergence without the traditional overhead of variational methods.

We trained three different models of size 7M, 37M, and 113M parameters. We found that both the small model and the medium model performed well and achieved the ground state after post-processing. The medium model is chosen as the best result as it performs best before optimization.

\begin{figure}[htbp]
\centering
\scalebox{0.90}{
\includegraphics[width=0.8\textwidth]{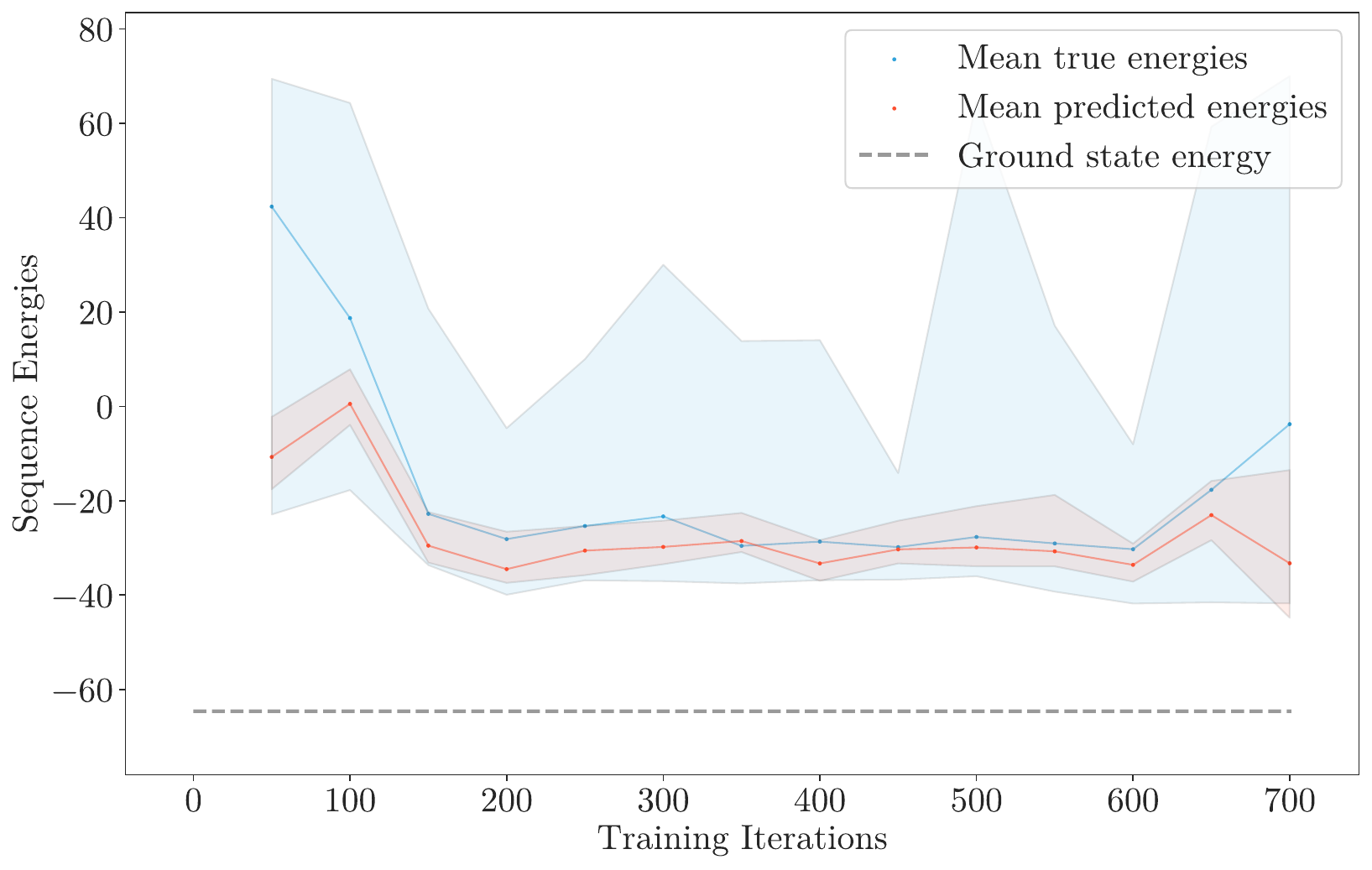}
}
\caption{Training convergence of the 113M parameter model on the Heisenberg model with $J=h=10$. The larger model fails to converge reliably due to insufficient training data, highlighting the importance of matching model capacity to available quantum evaluations.}
\label{fig:Fig3}
\end{figure}

The large model failed to converge, as shown in Fig.~\ref{fig:Fig3}, as the increase in parameters results in a need for more data. This requires an increase in QPU usage, which is computationally expensive.

\section{Operator pool} \label{app:C}

\begin{figure}[htbp]
\centering
\scalebox{0.90}{
\includegraphics[width=0.8\textwidth]{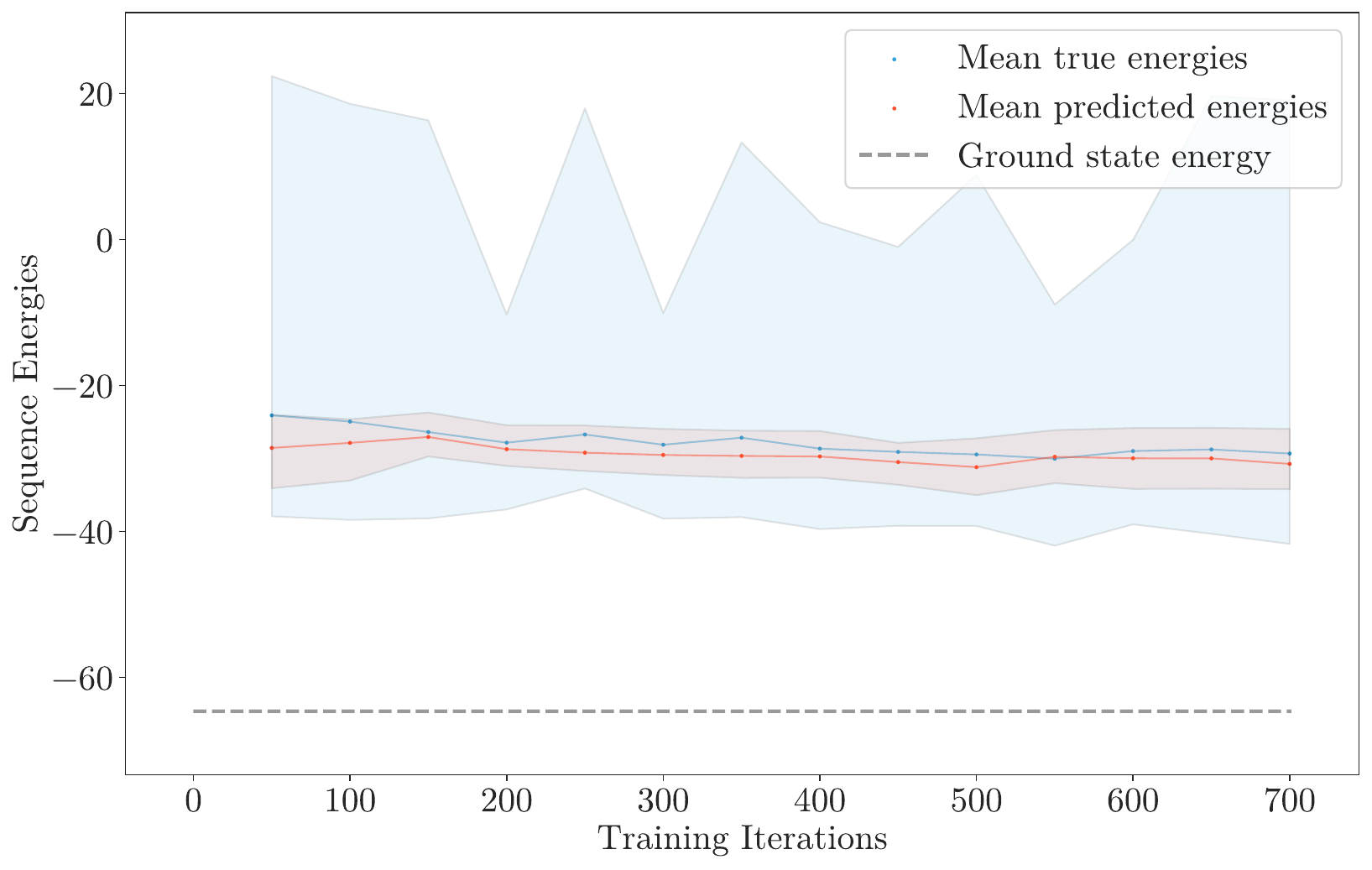}
}
\caption{Training convergence with an enlarged operator pool on the Heisenberg model with $J=h=10$. The expanded operator pool leads to less reliable convergence, demonstrating that a carefully constrained search space improves training efficiency for models with limited data.}
\label{fig:Fig6}
\end{figure}

We found that limiting the operator pool is essential for ground state convergence, especially for the lightweight models. To demonstrate this, we present results for training a model with an enlarged operator pool in Fig.~\ref{fig:Fig6}. This includes Pauli terms $\{X, Y, Z, XX, YY, ZZ\}$ as well as qubit pairs $\{i, i+2\}$ in addition to nearest-neighbor pairs.

\section{Choice of $\beta$ and $M$} \label{app:D}
\begin{figure}
    \centering
    \includegraphics[width=0.70\linewidth]{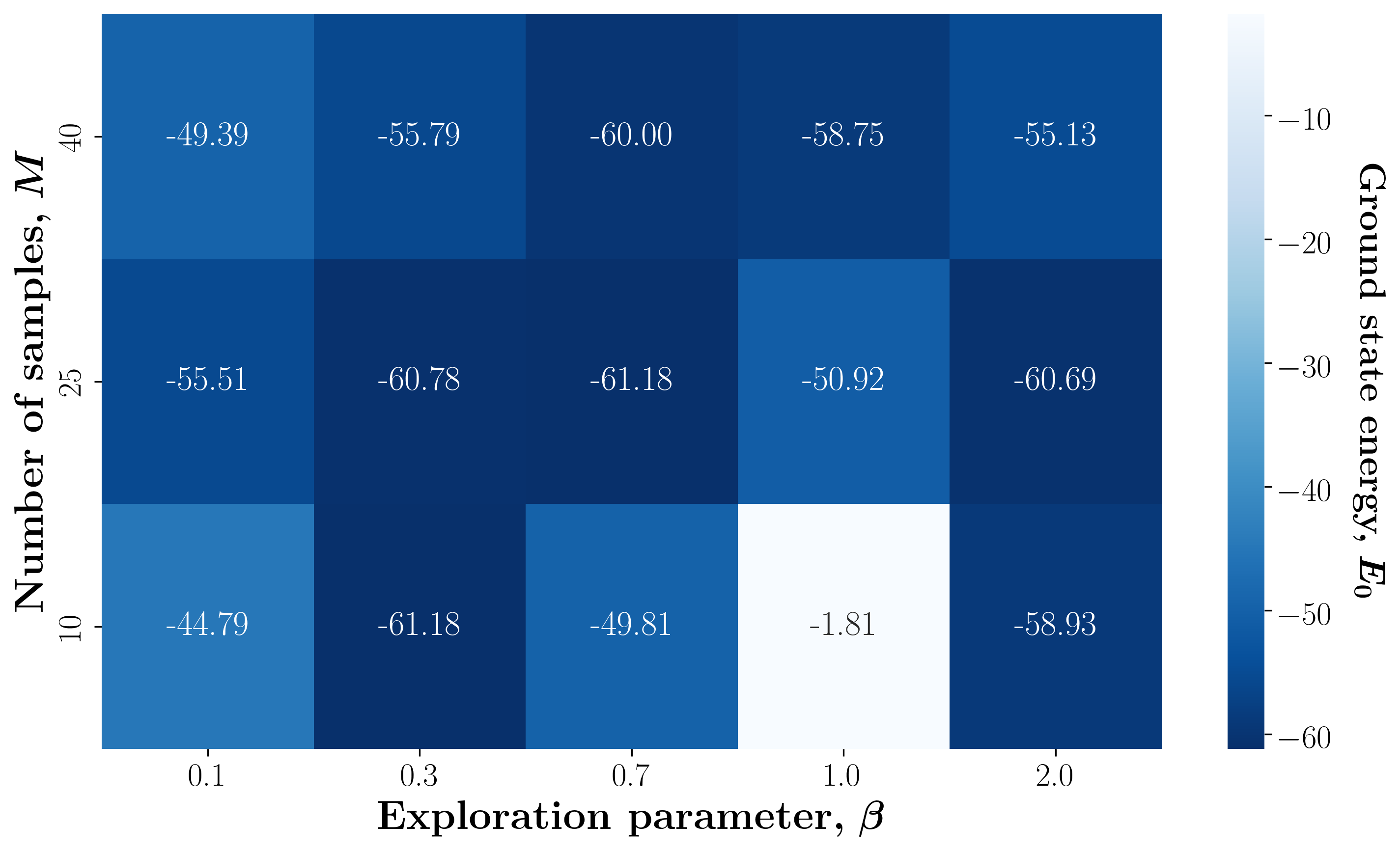}
    \caption{Ground state energy of the 1D Heisenberg Hamiltonian ($h=J=10$) as a function of the energy weighting parameter $\beta$ and circuits per epoch $M$, obtained from the 37M parameter model before applying post-processing optimization. The heatmap demonstrates that moderate values of $\beta$ and relatively small values of $M$ yield optimal performance.}
    \label{fig:gridsearch}
\end{figure}

We performed a grid search over the exploration parameter $\beta \in \{0.1, 0.3, 0.7, 1.0, 2.0\}$ and the number of circuits sampled at each epoch $M \in \{10, 25, 40\}$ for the one-dimensional Heisenberg Hamiltonian with $h = J = 10$. We trained the 37M parameter model for 700 epochs in each case and estimated the minimum ground state energy. The results are shown in Fig.~\ref{fig:gridsearch}. 

It is evident that sampling only $M=10$ circuits per epoch is adequate to reach a ground state energy as low as $-61.18$ J when $\beta$ is set to $0.3$. This demonstrates the sample efficiency of the approach, requiring relatively few quantum evaluations per training iteration.

A similar grid search for the case $J=1$, $h=10$ resulted in all combinations reaching the ground state energy $-37.0$ J, confirming the robustness of the method in the field-dominated regime.

\bibliographystyle{unsrtnat} 
\bibliography{reference} 

\end{document}